\documentclass[aps, twocolumn, prl, showpacs,preprintnumbers,amsmath,amssymb,superscriptaddress]{revtex4-1}
\usepackage{amsmath,amsfonts,amssymb,graphics,graphicx,epsfig,color,times,indentfirst,layout,hyperref,subfigure,multirow,bm}
\usepackage{hyperref}
\usepackage{ulem}
\hypersetup{linktocpage,,colorlinks=true,citecolor=red}

\usepackage{soul}

\newcommand{\be}{\begin{equation}}
\newcommand{\een}{\end{equation*}}
\newcommand{\bs}{\begin{split}}
\newcommand{\ben}{\begin{equation*}}
\newcommand{\ee}{\end{equation}}
\newcommand{\es}{\end{split}}
\newcommand{\bmx}{\begin{array}}
\newcommand{\emx}{\end{array}}
\newcommand{\bea}{\begin{eqnarray}}
\newcommand{\bean}{\begin{eqnarray*}}
\newcommand{\eea}{\end{eqnarray}}
\newcommand{\eean}{\end{eqnarray*}}
\newcommand{\dg}{^{\dagger}}
\newcommand{\dn}{^{\vphantom{\dagger}}}

\newcommand{\bp}{{\bf p}}
\newcommand{\lr}{\leftrightarrow}

\newcommand{\eps}{\epsilon}

\newcommand{\pref}[1]{(\ref{#1})}

\newcommand{\intinf}[1]{\int_{-\infty}^{+\infty}{#1}}
\newcommand{\intoinf}[1]{\int_{0}^{\infty}{#1}}

\newcommand{\intob}[1]{\int_{0}^{\beta}{#1}}

\newcommand{\re}[1]{{\rm Re}\left[ #1 \right]}
\newcommand{\im}[1]{{\rm Im}\left[ #1 \right]}
\newcommand{\tr}[1]{{\rm Tr}\Big[ #1 \Big]}

\newcommand{\abs}[1]{\left\vert #1 \right\vert}

\newcommand{\braket}[1]{\left\langle #1\right\rangle}

\newcommand{\bw}[1]{\begin{widetext}}
\newcommand{\ew}[1]{\end{widetext}}

\newcommand{\gray}[1]{}

\begin{document}
\title{Model for Ferromagnetic Quantum Critical Point in a 1D Kondo Lattice
}
\author{Yashar Komijani}
\affiliation{Department of Physics and Astronomy, Rutgers University, Piscataway, New Jersey, 08854, USA}

\author{Piers Coleman}
\affiliation{Department of Physics and Astronomy, Rutgers University, Piscataway, New Jersey, 08854, USA}
\affiliation{Department of Physics, Royal Holloway, University of
London, Egham, Surrey TW20 0EX, UK}
\date{\today}
\begin{abstract}
Motivated by recent experiments, we study a quasi-one dimensional
model of a Kondo lattice with Ferromagnetic coupling between the
spins. Using bosonization and dynamical large-$N$ techniques 
we establish the presence of a Fermi liquid and a magnetic phase separated by
a local quantum critical point, governed by the Kondo breakdown
picture. Thermodynamic properties are studied and a gapless charged
mode at the quantum critical point is highlighted.

\end{abstract}
\maketitle


Heavy fermion materials are a class of quantum system
in which 
the close competition between magnetism and intineracy drives a wealth of novel quantum ground
states, including hidden order, strange and quantum
critical metals, topological insulators and unconventional
superconductivity \cite{Wirth2016,Coleman2015}.  The various
entanglement mechanisms by which the localized magnetic moments
correlate and transform heavy fermion materials
provide an invaluable window on the governing 
principles needed to control and manipulate quantum matter. 

An aspect of particular interest is the 
quantum criticality that develops when a second-order magnetic 
phase transition is tuned to absolute zero.
In weakly interacting materials, magnetic quantum phase transitions
are understood in terms of the classic Slater-Stoner
instabilities of Fermi liquids (FLs), described by the interaction of soft
magnons with a Fermi surface according to the
Hertz-Millis-Moriya theory \cite{Hertz1976,Millis1993,Moriya1995}.  
The nature of quantum
criticality in strongly interacting materials, in which the magnetism
has a localized moment character, is less well
understood,  but is thought to involve a partial or complete
Mott localization of the electrons, manifested in heavy fermion compounds
as a break-down of the Kondo effect and a possible collapse
in the Fermi surface volume \cite{Si2001,Senthil2003,Coleman2010,Si2010}.   

Most research into heavy fermion quantum criticality 
has focused on antiferromagnetic instabilities, 
often discussed as a competition between the Kondo
screening of local moments, and antiferromagnetism, driven by 
the Ruderman-Kittel-Kasuya-Yosida (RKKY) interaction \cite{Doniach1977,Schroder2000,Loehneysen2007}. 
However, there is now a growing family
of heavy-fermion systems, including $\alpha$ and $\beta$-YbAlB$_4$  \cite{Nakatsuji2008,Matsumoto2011,Matsumoto2012}, 
YbNi$_4$P$_2$ \cite{Steppke2013}, YbNi$_3$Al$_9$   \cite{Baumbach2012} and CeRu$_2$Al$_2$B   \cite{Miyazaki2012}, in which the interplay of the Kondo effect and ferromagnetism is 
involved in the quantum criticality  \cite{Yamamoto2010}, {including engineered chains of quantum dots on metallic surfaces \cite{Lobos2012,Lobos2013} and itinerant systems \cite{Xu2017}.}

Motivated by these discoveries 
here we examine quantum criticality in a
Kondo lattice with ferromagnetic interactions. 
This affords many simplifications, for 
the uniform magnetization $M$ commutes with the Hamiltonian $[M,H]=0$ and is thus a conserved
quantity, free from quantum zero-point motion. 
Antiferromagnetic Kondo lattices are normally discussed in terms of a ``global'' phase diagram \cite{Si2001,Coleman2010} with two axes - 
the Doniach parameter $x=T_{K}/J_{H}$, set by the ratio of the Kondo temperature $T_{K}$ to the Heisenberg coupling $J_{H}$, and the frustration parameter $y$ 
measuring the strength of magnetic zero
point fluctuations. The elimination of magnetic zero point fluctuations
allows us to focus purely on the $x$-axis of the global phase
diagram. Moreover, it now 
becomes possible to study magnetic quantum criticality in a one dimensional
model. 

\begin{figure}[t] \includegraphics[width=\linewidth]{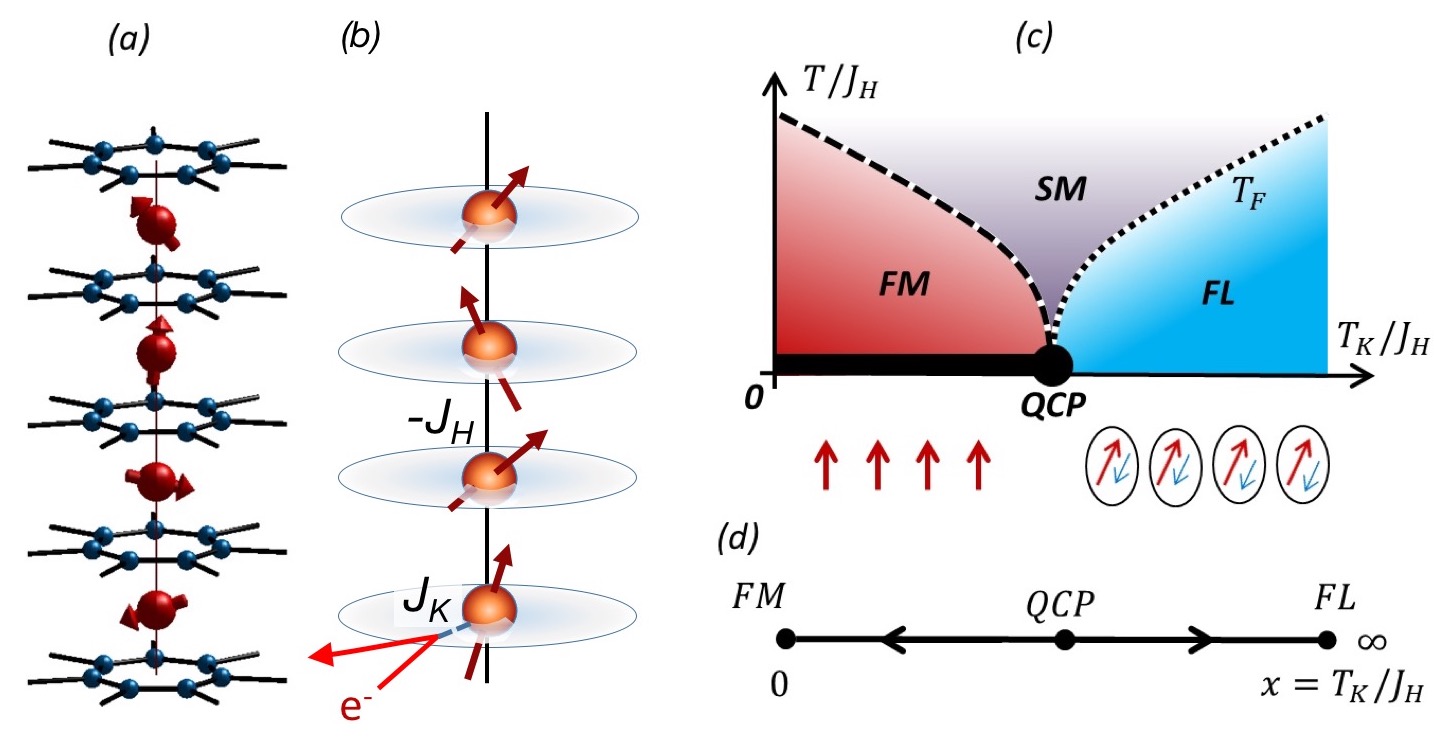}
\caption{(a) The quasi-1D structure of Yb local
moments (red) in YbAlB$_4$ sandwiched between conducting B layers. 
(b) 1D model, showing local moments (orange),  coupled via a ferromagnetic
coupling $-J_H<0$, each Kondo-coupled to a separate conduction electron sea (white-blue
layers). (c) Phase diagram we find for the model as a
function of $T_K/J_H$ and temperature, showing Fermi liquid
and 1D ferromagnetic regime, 
separated by a QCP. The Fermi temperature of the FL, $T_F$ vanishes
at the QCP. 
The 1D FM only orders at zero temperature and is intrinsically quantum critical.
(d) RG flow of transverse Ising model to which our model maps in the Ising limit.}\label{fig1}
\end{figure}


Our model is motivated by the quasi-one dimensional Yb structure  of 
YbAlB$_{4}$, in which a chain of ferromagnetically coupled Yb spins
hybridizes with multiple conducting planes of B atoms
(Fig. 1.) \cite{Nevidomskyy2009}.   For
simplicity, we treat
each plane as an autonomous electron bath, individually
coupled via an antiferromagnetic Kondo coupling $J_{K}$, according to
\begin{equation}\label{}
H = \sum_j
\biggl (H_{c} (j) + J_{K} \vec{S}_{j}\cdot \vec{\sigma }_{j} - J_{H}\vec{S}_{j}\cdot\vec{S}_{j+1}\biggr),
\end{equation}
where  $\vec{S}_{j}$ is the spin at the $j$-th site, 
coupled ferromagnetically to its neigbor with strength $J_{H}$. 
$H_c (j) =\sum_{{\bf p}}\eps_{\bp } c\dg_{\bp\alpha} (j)
c_{\bp \alpha} (j)$ describes the $j-th$ layer of electrons, coupled
to the chain via its spin density
$\vec\sigma_{j} = \psi \dg_{j\alpha } \vec{\sigma }_{\alpha \beta
}\psi_{j\beta } $ at the chain, where $\bp$ is the momentum of the conduction electrons at the j-th layer and $\psi\dg _{j\alpha } = 
\sum_{ \bp }c\dg _{\bp \alpha } (j)$ creates an electron at the position of impurity site on the chain.

At small $x=T_K/J_H$ the 1D chain is ferromagnetically correlated, 
developing true long range order only at zero temperature, 
while at large $x$ it forms a paramagnet where each
spin is individually screened: in
between, there is a  quantum critical point (QCP) \cite{Doniach1977,Yamamoto2010}.
This QCP has been demonstrated \cite{Lobos2013} in the Ising limit of this Kondo lattice at the Toulouse decoupling point \cite{toulouse69,emery92}, 
which permits bosonization of the Hamiltonian,  mapping it \cite{Komijani2017} onto 
the transverse field Ising model, $H\to T_K\sum_n S_n^x+J_H^z\sum_n
S_n^zS_{n+1}^z$. This model has a well-known  RG flow [Fig.\ref{fig1}(d)]
and a quantum phase transition at $J_H^z=T_K$  \cite{Sachdev2011}.
However, in this limit, the stable phases are gapped
and to gain a deeper insight into the physics of the QCP,  
we return to the Heisenberg limit. 

Here instead, we use a large-$N$ Schwinger boson approach which treats
the magnetism in the Heisenberg limit, while also explicitly preserving the Kondo effect. 
Our method 
unifies the Arovas and Auerbach treatment of ferromagnetism \cite{Arovas1988} with the description
of the Kondo problem by Parcollet, Georges et al
 \cite{Parcollet1997,Parcollet1998,Coleman2005,Rech2006}. 
An important aspects of this approach, is the use of a multi-channel
Kondo lattice in which the spin $S$ and the number
of channels $K$ is commensurate ($K=2S$), allowing for a perflectly
screened Kondo effect \cite{Rech2006}. 

Figure\,1(c) summarizes the key results. 
At large $T_{K}/J_{H}$ our method describes a FL
phase with Pauli susceptibility $\chi\sim 1/T_F$ and a linear specific
heat coefficient $\gamma= C/T \sim 1/T_F $. As $x$ is reduced to a
critical value $x_{c}$, the characteristic scale $T_{F} (x)$,
determined from the  magnetic susceptibility and linear specific
heat coefficient (Fig. 2 (c,d)), 
drops continously to
zero, terminating at a QCP. 
This suppression of $T_{F}$ resembles the
Schrieffer mechanism for the reduction of the Kondo temperature
in Hund's metals \cite{Schrieffer1967,Larkin1972,
Haule2009,Nevidomskyy2009a}. The large $N$  QCP is characterized by
powerlaw dependences of the specific heat, local and uniform
susceptibilities. 
\begin{equation}\label{}
\chi (T)\sim \frac{1}{T}, \qquad \chi_{loc}(T)\sim\frac{1}{T^{1-\alpha}},\qquad \frac{C}{T}\sim \frac{1}{{T^{\alpha }}}
\end{equation}
where the exponent $\alpha[s]<1$ 
is  function of the spin  $s=2S/N$. 
At still smaller $x$ the chain develops a
fragile Ferromagnetism which disappears at finite
temperatures.  Here $\chi\sim 1/T^2$ and $C/T\sim
1/\sqrt{T}$ characteristics of a critical 1D Ferromagnetism (FM).
There are two notable aspects of the physics:
first, the QCP exhibits an emergent critical 
charge fluctuation mode associated with Kondo breakdown, and secondly 
the 1D ferromagnetic ground-state is intrinsically quantum critical, 
transforming into a Fermi liquid with characteristic scale  of order
the Zeeman coupling, upon application of a magnetic field.  This last feature
is strongly reminiscent of the observed physics of $\beta
-$YbAlB$_{4}$, a point we return to later.

Our large $N$ approach is obtained by casting the local
moments as Schwinger bosons
$S (j)_{\alpha\beta}=b\dg_{j\alpha}b\dn_{j\beta}$, where 
$2S = n_{b} (j)$ is the number of bosons per site, each individually
coupled to a $K$ channel conduction sea, with Hamiltonian
\begin{equation}\label{}
H=\sum_j[
H_{FM} (j) +H_{K} (j) +H_{C} (j)+\lambda_{j} (n_{b} (j)-2S)],
\end{equation}
where (scaling down coupling constants)
\bea
H_{FM} (j)&=&- (J_H/N)( b\dg_{j\alpha}b\dn_{j+1,\alpha})( b\dg_{j+1,\beta}b\dn_{j\beta})
\nonumber\\
H_K (j)&=&-(J_K/N)\bigl (b\dg_{j\alpha}\psi \dn_{ja\alpha} \bigr )
(\psi\dg_{ja\beta}b\dn_{j\beta})
\nonumber\\
H_C (j)&=&\sum_{\bp }\eps_{\bp } c\dg_{\bp a \alpha } (j) c\dn_{\bp a\alpha } (j)\label{eq3},
\eea
where $\lambda_{j}$ is a Lagrange multiplier that 
imposes the 
constraint.
Here we have adopted a summation convention, with implicit summations
over the (greek ) $\alpha\in[1,N]$ spin and (roman) $a\in [1,K]$ channel
indices. In the calculations, we take $2S=K=sN$ for perfect screening,
where $s$ is kept fixed.

Next, 
we carry out the Hubbard-Stratonovich
transformations:
\bea
H_K (j)&\to&
\bigl [
(b\dg_{j\alpha}
 \psi \dn_{ja\alpha})
\chi_{ja} 
+{\rm h.c}\bigr ]+\frac{N\bar\chi_{ja}\chi\dn_{ja}}{J_K}\\
H_{FM} (j)&\to&
\bigl [\bar  \Delta_{j}(b\dg_{j+1,\alpha} b\dn_{j,\alpha})+{\rm h.c}\bigr ]
+\frac{N  |\Delta_{j}|^{2}}{J_H}\nonumber.
\eea
The first line is the Parcollet-Georges factorization 
of the Kondo interaction, where the $\chi_{ja}$ are charged, spinless
Grassman fields that mediate the Kondo effect
in channel
$a$. The second line is the Arovas-Auerbach factorization of
the magnetic interaction in terms of the 
bond variables $\Delta_{j}$ 
describing the spinon delocalization. 
Both $b$ and $\chi$ fields have non-trivial dynamics
\cite{Parcollet1997,Parcollet1998,Coleman2005,Rech2006}, with self-energies 
given by \cite{SM}
\be
\Sigma_\chi(\tau)=g_0(-\tau)G_B(\tau), \quad
\Sigma_B(\tau)=-k g_0(\tau)G_\chi(\tau).\label{eqself}
\ee
Here $G_{\chi } (\tau )$, $G_{B} (\tau )$ and $g (\tau )$ are the
local propagators of the holons, spinons and conduction electrons,
respectively. 
The conduction electron self-energy
is 
of order
$O(1/N)$ and is neglected in the large-$N$ limit, so that
$g_{0}(\tau)$is the bare local conduction
electron propagagator.
The holon Green's function 
is purely local, given by 
$G_\chi (z)=[{-J^{-1}-\Sigma_\chi(z)}]^{-1}$,
but the interesting new feature of our calculation 
is the delocalization of the spinons along the chain.
Seeking uniform solutions where  $\Delta_{j}= -\Delta$ and $\lambda_{j}=\lambda
$, the spinons develop a 
dispersion $\eps_{B} (p) =-2\Delta\cos p$, with 
propagator 
$G_{B} (p,z)= [z - \eps_{B} (p) - \lambda - \Sigma_{B} (z)]^{-1}$. The
momentum-summed 
{\it local} propagator is then 
\be
G_{B}(z)=\sum_p
G_B(p,z)=\int{\frac{d\eps_{B}\rho(\eps_{B})}{z-\lambda-\eps_B-\Sigma_B(z)}}\label{eqG}
\ee
where $\rho(\eps_B)=(2\pi\Delta)^{-1}[1-(\eps_B/2\Delta)^2]^{-1/2}$
is the bare spinon density of states.  Using Cauchy's theorem, 
\be
G_B(z)=\frac{1}{\Omega[z]}\frac{1}{\sqrt{1-[\Omega(z)/2\Delta]^{-2}}}\label{eqGB}
\ee
where $\Omega(z)\equiv z-\lambda-\Sigma_B(z)$ \cite{SM}.

Stationarity of the Free energy with respect to $\lambda$ and $\Delta
$ then leads to two saddle-point equations 
\be
\intinf{\frac{d\omega}{\pi}n_B(\omega)\im{G_B(\omega-i\eta)}}=s,\label{eqcons}
\ee
\begin{equation}\label{eqJH}
\frac{1+\zeta
\frac{\Delta^{2}}{J_{H}^{2}} }{J_{H}} =
\int{\frac{d\omega}{2\pi \Delta^{2} }}n_B(\omega)\im{\Omega(z)G_B(z)}_{z=\omega+i\eta }
\end{equation}
which determine $\lambda$ and $J_{H}$ self-consistently. 

In (\ref{eqJH}) we have added an  additional $\zeta
\frac{\Delta^{2}}{J_{H}^{2}}$ which stabilizes the quantum critical point. 
Schwinger boson mean-field theories suffer 
from weak first order phase transitions upon development of 
finite $\Delta $, due to fluctuation-induced  attractive
quartic $O (\Delta^{4})$  terms in the effective
action. This difficulty \cite{DeSilva02}, 
has thwarted the study of quantum criticality with this method.
These first order transitions are actually 
a non-universal 
artifact of the 
way the large $N$ limit is taken, easily circumvented 
by adding a small repulsive  biquadratic term $H' (j) = \zeta J_{H}
(\vec S_j\cdot\vec S_{j+1})^2$ to the Hamiltonian. 
For an $SU(2)$ $S=1/2$ moment, the biquadratic term can be absorbed into the
Heisenberg interaction, 
but for the higher spin representations of the large $N$ expansion,
it contributes a positive
quartic correction $O (\zeta \Delta^{4})$ to the effective
action that restores the second-order phase transitions (at both zero and
finite temperature) to the large $N$ limit \cite{SM}.
In practice, a $\zeta\sim 0.001$ is sufficient to remove the first order transition, so that $\Delta $ tunes linearly with $J_{H}$ across the quantum critical
point.

To find $G_B(\omega)$ and $G_\chi(\omega)$  we solve
Eqs.\,(\ref{eqself}-\ref{eqcons}) self-consistently on a linear and
logarithmic grid. 
The entropy formula from \cite{Rech2006,Coleman2005} was used to compute the
specific heat associated with these solutions \cite{SM}. 

\begin{figure}[t!]
\includegraphics[width=\linewidth]{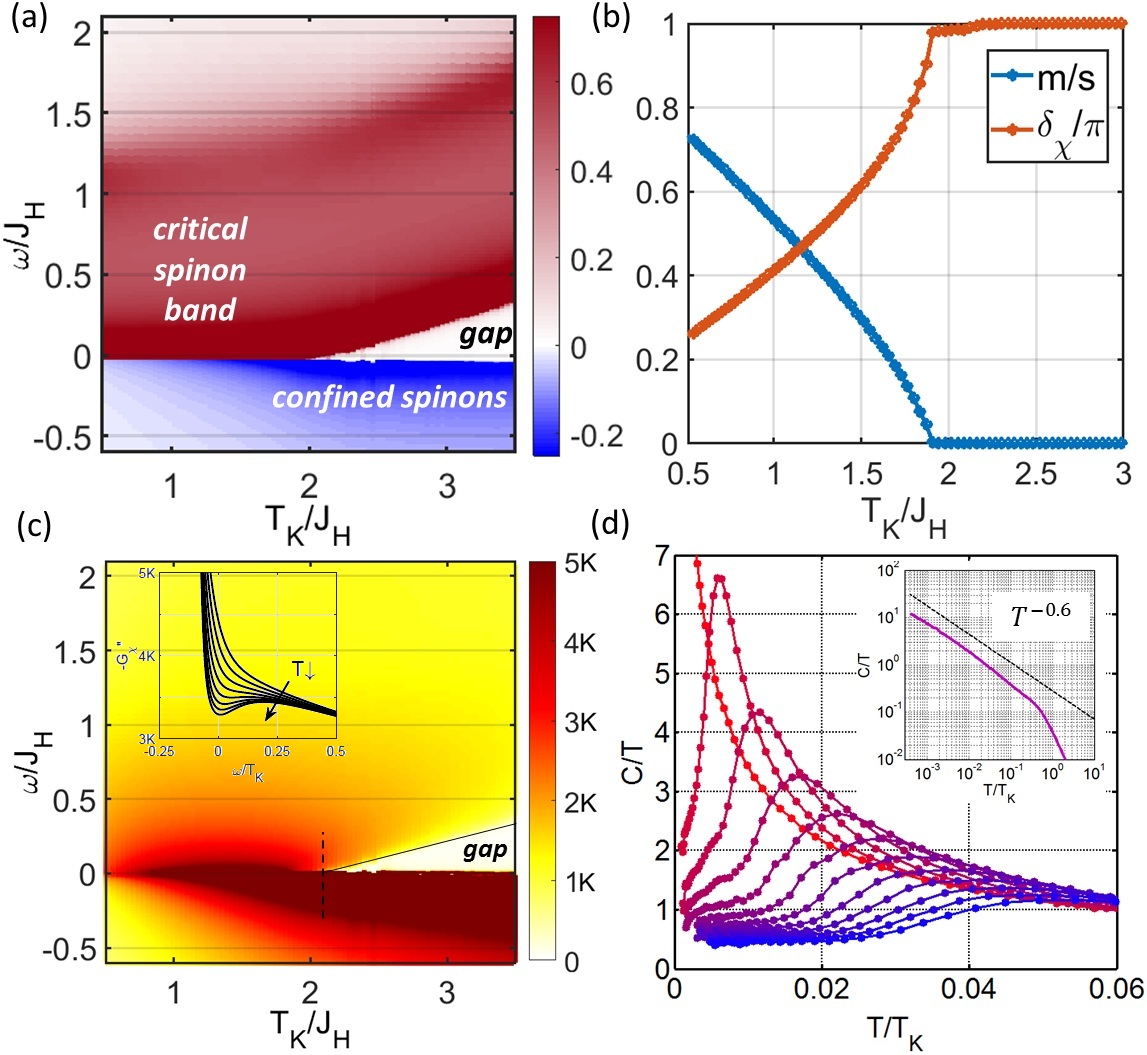}
\caption{(a) The spectral density of spinons $-G''_B(\omega+i\eta)$ for
$k=s=0.3$ as a function of $T_K/J_H$, shows spinon band at positive energy and the Kondo-screened spins appearing as confined spinons at negative energy. The Kondo gap at large $x$, shrinks linearly with loweing $x$,
collapsing at about $x\approx 2$. (b) Zero temperature
magnetization $m/s$ (blue) and holon phase shift $\delta_\chi/\pi$ (red) as a function of $T_K/J_H$.
(c) The spectral density of holons $-G''_\chi(\omega+i\eta)$ as a function of $T_K/J_H$ shows the Kondo-gap collapse and the critical mode at the QCP (inset). (d) specific heat
coefficient $\gamma(T)={C/T}$ vs. temperature as $T_K/J_H$ is varied from 5
(blue) to 0.1 (red). 
The inset in (d) shows the power law dependence of $\gamma$ at the
QCP.
}\label{fig2}
\end{figure}

In the Kondo limit ($T_{K}/J_{H}\gg 1$) 
the local moments are fully 
screened, forming a Fermi liquid; in the Schwinger boson scheme, 
the formation of Kondo-singlets 
is manifested as a spectral gap $\Delta_{g}\sim T_{K}$ \cite{Rech2006} in the
spectrum of the spinons and holons, where $T_K=f(T_K^0,s)$ and $T_K^0=De^{-1/\rho J}$ is
the Kondo temperature (Fig.\,\ref{fig2}(a)). The opening of this gap effectively confines the spinon and conduction electron into a singlet bound state, leaving behind an
elastic resonant scattering potential which satisfies the Friedel sum rule
with phase shift $\delta = \pi/N$. 

In the opposite ferromagnetic limit $T_{K}/J_{H}\ll 1$, the chain forms
a fragile ferromagnet. In this case, 
the spinons are condensed in the ground-state, but 
at finite temperatures, the spinon band is gapped: 
the constraint
\pref{eqcons} ensures that the gap in the spectrum grows quadratically, 
$\Delta_b(T)\propto T^{2}$, and together with the quadratic dispersion,
this leads to a free energy $F\propto
T^{3/2}$, a critical susceptibility $\chi\propto
T^{-2}$ and a specific heat coefficient $C/T \propto T^{-1/2}$
\cite{Arovas1988,SM}, 
in agreement with Bethe ansatz
\cite{Schlottmann1985,Schlottmann1986,Takahashi1985,Takahashi1987}
. The van-Hove singularity of density of states means that the ferromagnet is fragile, so that 
the bosons only condense, developing true long-range order 
at absolute zero.

Fig.\ref{fig2} shows the evolution of properties between these two limits. 
As $x$ is reduced, 
the spectral gap responsible for Fermi liquid behavior
shrinks linearly to zero 
at the QCP at $x_{c}\approx 2$, an 
indication of Kondo break-down. 
This suppression of the Kondo temperature with $x$ is closely analagous
to reduction of the Kondo temperature by Hund's
coupling \cite{Schrieffer1967,Okada1973,Nevidomskyy2009a}, with $\Delta
\sim J_{H}$ playing the role of the Hund's coupling 
and the ratio $\xi/a$ of the spin correlation length to the lattice
spacing, playing the role of the effective moment. 

The ground-state ferromagnetic moment is given by
\begin{equation}\label{}
m= {\lim_{T\rightarrow 0}}\int_{0}^{\infty }\frac{d\omega}{\pi  }n_{B}(\omega) {\rm Im}G_{B}
(\omega-i\eta ).
\end{equation}
which measures the residual positive-energy spinon
population, which condenses at $T=0$ (Fig.\,\ref{fig2}(b)). $m$ is zero in the fully screened state, and
rises gradually to a maximum value $m = s =2S/N$ 
in the ferromagnetic limit.   
Note that $m/s<1$ indicates that the magnetic moment is
partially screened by an incipient Kondo effect which continues into
the fragile magnetic phase. 

Although our simple model does not allow us to examine the evolution
of the Fermi surface, we can monitor the delocalization of heavy electrons
by examining the phase shift of the holons $\delta_{\chi } = {\rm
Im}\ln [-G_{\chi }^{-1} (0-i\delta )]$.  
The change in the number of delocalized heavy electrons $\Delta
n_{f}$ is related to the holon phase shift by the relation 
$\Delta n_{f}=\sum_a (\frac{\delta_{\chi }}{\pi})$ \cite{Coleman2005,Rech2006}, which is plotted as a function of $x$ in Fig.\,\ref{fig2}(b).
Although we do not observe a jump in $\Delta n_f$ at
the QCP, there is a sharp cusp in its evolution at $x=x_{c}$. One of the interesting aspects of our results, is that the
holon spectrum becomes critical at the QCP (Fig.\,\ref{fig2}(c), inset), signaling the emergence of
a critical spinless charge fluctuation that accompanies the 
critical formation and destruction of singlets.

The specific heat coefficient
$\gamma\equiv C/T=dS/dT$, plotted in Fig.\,\ref{fig2}(d) shows a
``Schottky'' peak at $T\sim T_F$ for large $x$ (blue) which
collapses to zero as $x\rightarrow x_{c}$(red). 
At the QCP, $\gamma (T)\sim T^{-\alpha }$
follows a power-law, where $\alpha [s]$ depends on the reduced spin
$s=2S/N$. In the calculations displayed here, $\alpha =0.6$ for
$s=0.3$ (Fig. \ref{fig2} d).
In the magnetic phase
$\gamma\sim 1/\sqrt{T}$ again characteristic of 1D FM. 

\begin{figure}[h!]
\includegraphics[width=1\linewidth]{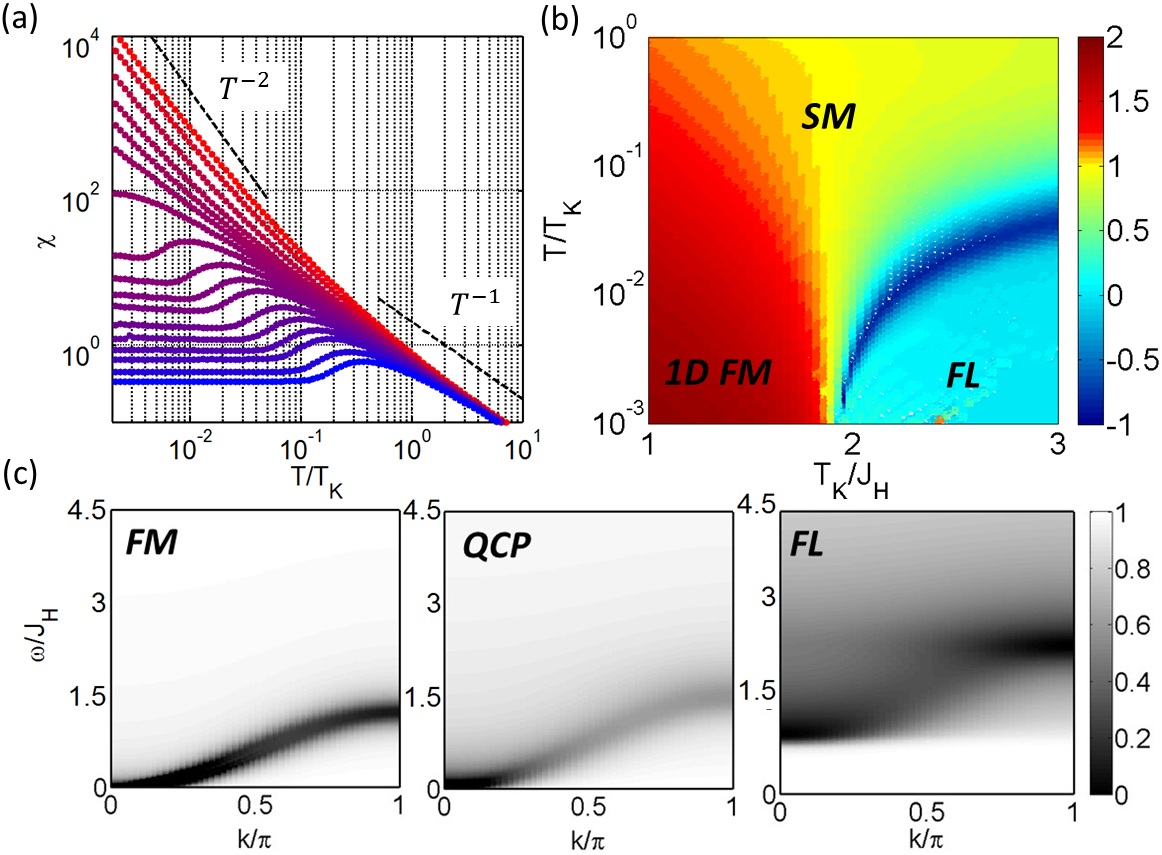}
\caption{(a) Uniform spin susceptibility $\chi$ as a function of temperature as $T_K/J_H$ is varied from 0.1 (red) to 5 (blue). (b) The phase diagram obtained from the temperature-exponent $\kappa$ of susceptibility $\chi\sim T^{-\kappa}$, shows the Kondo breakdown induced by the Schrieffer suppression of the Fermi temperature and separated from the magnetic phase by a QCP. (c) Dynamical spin susceptibility in FL ($T_K/J_H=3.6$), QCP (1.65) and FM (0.36) regimes, respectively.}\label{fig3}
\end{figure}


Fig.\,\ref{fig3}(a) shows the dependence of the 
uniform spin susceptibility on $x$. In the Fermi liquid at 
large $x$ (blue), there is a cross over from 
a Curie susceptibility $\chi\sim 1/T$ at high-$T$ 
to a Pauli susceptibility $\chi\sim 1/T_F$ at the Fermi temperature $T_{F}$.
As $x$ decreases, $T_F$ decreases to zero and the
susceptibility becomes critical.  At the QCP the susceptibility
$\chi \sim 1/T$ follows a simple Curie law.  For $x<x_{c}$, the susceptibility
displays a $\chi\sim 1/T^2$
characteristic of 1D FM. We use the dependence of the temperature
exponent 
$\kappa=-d\log\chi/d\log T$ of the 
susceptibility on $x$ and temperature to map out the phase diagram (Fig.\,\ref{fig3}(b)). The dark blue stripe delineates the renormalized Fermi temperature 
of the Fermi liquid, showing its collapse to zero as $x\rightarrow
x_{c}^{+}$. The corresponding evolution in the 
dynamical magnetic  susceptibility  $\chi '' (q,\omega)$
of various phases are shown in Fig.\,\ref{fig3}(b). The sharp magnon band in the magnetic phase is smeared at the QCP, denoting fractionalization of the spins. The FL phase features a spectral gap, which is an artificant of large-$N$ method, as well as some remnants of the magnon band.


{We have also studied the effect of a magnetic field
\cite{SM}. 
While the Fermi liquid is robust, 
application of a small magnetic field to the QCP or the FM phase
\cite{Coleman2003,Coleman2005a} immediately reinstates 
Fermi-liquid behavior with an scale $T_B$ set by the Zeeman energy 
(at the QCP) or a combination of the spinon
bandwidth and magnetic field (in the FM phase) \cite{SM}. 
The ferromagnetic phase is thus intrinsically quantum critical. 
}


There are two interesting possibilities arising from our work; 
The intrinsic quantum criticality of the 1D FM phase is reminiscent of
$\beta-$YbAlB$_{4}$. 
This raises the fascinating question
as to whether the critical FM seen in our model 
might be stabilized by frustration, as a phase
in higher dimensions. 
Second concerns the character of the Kondo break-down at the
QCP, which  appears to involve a critical spinless charge
degree of freedom. It
is intriguing to speculate whether this might be an essential element
of a future theory of heavy fermion quantum criticality. 

A natural extension of our work is the 
anti-ferromagnetism, which allows an exploration of the  
global phase diagram. 
Generalizations to
higher dimensional systems can be envisioned by using our approach 
as an impurity or cluster solver for dynamical mean-field theory \cite{Georges1996}.


\section*{Acknowledgement}
 
This work was supported by a Rutgers University Materials Theory
postdoctoral fellowhsip (YK), and by NSF 
grant DMR-1309929 (PC).   We gratefully acknowledge discussions with
Thomas ~Ayral, Collin Broholm, Elio ~K\"onig, Satoru Nakatsuji, 
and Pedro ~Schlottmann.



\bibliography{FQCP} 
\newpage
\section{Supplementary material}
\subsection{Ferromagnetically coupled spin chain}
Here, we briefly review the Arovas and Auerbach \cite{Arovas1988} treatment of 1D ferromagnetism.
In absence of Kondo coupling, the Hamiltonian is $H_\lambda+H_{FM}$ where $H_\lambda$ refers to the Lagrange multipliers. We assume a uniform mean-field solution $\Delta$ and find the temperature-dependence of the chemical potential. 

{\it Chemical potential} - The number of bosons satisfies
\be
s=\int_{-\pi}^{\pi}{\frac{dk}{2\pi}}n_B(\lambda+\eps_k)=\int{d\eps}\rho(\eps)n_B(\eps+\lambda),
\ee
When $T\ll\Delta$, the Bose Einstein function $n_{B}$ is highly
focused at low energies, and the physics is dominated 
by the quadratic bottom of the spinon band where $\eps_k=-2\Delta\cos k\approx -2\Delta+\Delta k^2$ so that
\be
\rho(\eps)=\frac{2}{2\pi d\eps_k/dk}=\frac{1}{2\pi\Delta\sin k}\sim\frac{1}{2\pi\sqrt{\Delta}\sqrt{\eps+2\Delta}}.
\ee
The factor of $2$ in the numerator derives 
from the 2-to-1 relation between the momentum and energy $\eps_k$.
Using $\eps'=\eps+2\Delta$ and $\lambda'=\lambda-2\Delta$, we can write
\bea
s&=&\frac{1}{2\pi\sqrt{\Delta}}\intoinf{d\eps'}\frac{1}{\sqrt{\eps'}}\frac{1}{e^{\beta(\lambda'+\eps')}-1}\nonumber\\
&=&\frac{1}{2\pi}\sqrt{\frac T\Delta}\Gamma(1/2){\rm Li}_{1/2}(z),\label{eqq}
\eea
where we have defined the fugacity $z\equiv e^{-\beta\lambda'}$ and used that
\bea
I_p(z)&\equiv&
\intoinf{dx\frac{x^{p-1}}{e^x/z-1}}\nonumber\\
&=&\sum_{k=0}^\infty \frac{z^{k+1}}{(k+1)^{p}}\intoinf{dy}y^{p-1} e^{-y}\nonumber\\
&=&\Gamma(p){\rm Li}_p(z),\label{eqIp}
\eea
in terms of polylogarithm function ${\rm Li}_p(z)$. For $z=1$ and $p>1$ the series is convergent and ${\rm Li}_p(z)=\zeta(p)$ where $\zeta(p)$ is the Riemann zeta function. But for $p<1$ we also get convergence if $\abs{z}<1$. 
Due to the $\sqrt{T}$ prefactor in the expression for  $s$ in Eq.\,\pref{eqq}, the function ${\rm Li}_{1/2}(z)$ has to diverge  as $T\to 0$ so that $s$ stays constant and this happens for $z\to 1$ for which we have
\be
\lim_{z\to 1}{\rm Li}_{1/2}(z)=\frac{\sqrt{\pi}}{\sqrt{-\log(z)}}=\frac{\sqrt{\pi}}{\sqrt{\lambda'/T}}.
\ee
Therefore, we conclude
\be
{\lambda'=\alpha T^2}, \qquad \alpha=\frac{\Gamma^2(1/2)}{2\pi \Delta s^2}.\label{eqalpha}
\ee

{\it $\Delta$ vs. $J_H$ - } This can also be alternatively written as
\be
\frac{\Delta}{J_H}=-\frac{1}{2\Delta}\int{d\eps}\eps\rho(\eps)n_B(\lambda+\eps)
\ee
Again assuming that the bottom of the band is only involved (this works in the limit of large $\Delta/T$) we have
\bea
2\frac{\Delta^2}{J_H}&=&-\frac{1}{2\pi\sqrt{\Delta}}\intoinf{d\eps'}\frac{\eps'-2\Delta}{\sqrt{\eps'}}\frac{1}{e^{\beta(\eps'+\lambda')}-1}\\
&=&-\frac{1}{2\pi\sqrt{\Delta}}\Big[T\sqrt{T}I_{3/2}(z)-2\Delta \sqrt{T}I_{1/2}(z)\Big],
\eea
in terms of $I_p(z)$ defined in Eq.\,\pref{eqIp}. Equivalently
\be
\frac{2\Delta^{5/2}}{J_H}=-\frac{1}{2\pi}\Big[T\sqrt{T}\Gamma(3/2){\rm Li}_{3/2}(z)-2\sqrt{T}\Delta\Gamma(1/2){\rm Li}_{1/2}(z)\Big].
\ee
So, using $z\partial_z{\rm Li}_p(z)={\rm Li}_{p-1}(z)$ we find 
\be
2\frac{\Delta^{5/2}}{J_H}=\frac{1}{2\pi}\Big[2\Gamma(1/2)\times \sqrt{\pi}\frac{\Delta}{\sqrt{\alpha_\Delta}}+T^2\sqrt{\alpha_\Delta}2\sqrt\pi\Gamma(3/2)\Big].
\ee
At zero temperature, we can drop the second term in the right side and from $\alpha_\Delta\propto s^2/\Delta$ find
\be
2\frac{\Delta^{5/2}}{J_H}=2s\Delta^{3/2}\quad\rightarrow\quad \Delta(T\to 0)= J_Hs.
\ee
{\it Susceptibility -} The susceptibility is
\bea
\chi=-\frac{dM}{dB}\Big\vert_{B=0}&=&\frac{\beta}{L}\sum_kn_B(\lambda+\eps_k)[1+n_B(\lambda+\eps_k)]\nonumber\\
&=&
-\frac{1}{L}\sum_k \frac{-\beta e^{\beta(\lambda+\eps_k)}}{(e^{\beta(\lambda+\eps_k)}-1)^2}
\eea
Doing the momentum sum we find
\bea
\chi&=&\beta\int\frac{dk}{2\pi}\frac{e^{\beta(\lambda+\eps_k)}}{[e^{\beta(\lambda+\eps_k)}-1]^2}\nonumber\\
&=&\beta\intoinf{d\eps'}\rho(\eps'-2\Delta)\frac{e^{T\eps'}/z}{[e^{T\eps'}/z-1]^2}\nonumber\\
&=&\frac{\beta}{4\pi\sqrt{\Delta}}\sqrt{T}\intoinf dx\frac{1}{\sqrt{x}}\Big\{\frac{e^{x}/z}{[e^{x}/z-1]^2}\Big\}\nonumber\\
&=&\frac{T^{-1/2}}{4\pi\sqrt{\Delta}}\Gamma(1/2){\rm Li}_{-1/2}(z)\nonumber\\
&\to&\frac{\alpha^{-3/2}}{8\sqrt{\pi}\sqrt\Delta}\frac{1}{T^2}.
\eea
where the last line is valid in the limit of low-temperature.\\

\noindent{\it Free energy} - This is simply
\bea
F+\lambda s&=&\int{\frac{d\omega}{\pi}}n_B(\omega)\int{d\eps}\rho(\eps)\im{\log(\eps+\lambda-\omega-i\eta)}\nonumber\\
&=&\int{\frac{d\omega}{\pi}}\frac{n_B(\omega)}{2\pi\sqrt{\Delta}}\int_{0}^\infty{\frac{d\eps'}{\sqrt{\eps'}}}\im{\log(\eps'+\lambda'-\omega-i\eta)}\nonumber\\
&=&-\pi\frac{1}{2\pi\sqrt{\Delta}}\int{\frac{d\omega}{\pi}}n_B(\omega)\int_{0}^{\omega-\lambda'}{\frac{d\tilde\eps}{\sqrt{\tilde\eps}}}\nonumber\\
&=&-\pi\frac{1}{\pi\sqrt{\Delta}}\int{\frac{d\omega}{\pi}}n_B(\omega)\sqrt{\omega-\lambda'}.
\eea
After a $\omega=\lambda'+Tx$ change of variable,
\bea
F&=&-\lambda s-\frac{1}{\pi\Delta}T^{3/2}\intoinf{\frac{dx\sqrt{x}}{e^x/z-1}}\nonumber\\
&=&-2\Delta s-\lambda' s-\frac{T^{3/2}}{\pi\Delta}\Gamma(3/2){\rm Li}_{3/2}(z)\label{eqFFM}
\eea
The polylogarithm has the expansion
\be
{\rm Li}_{3/2}(z)=-2\sqrt{\pi\log(-z)}+\sum_{m=0}\frac{\log^m(z)\zeta(1/2-m)}{m!}.\nonumber
\ee
Inserting this and also Eq.\,\pref{eqalpha} into Eq.\,\pref{eqFFM},
\bea
F&=&-2\Delta s-\frac{T^2\Gamma(1/2)}{2\pi\Delta s}[\Gamma(1/2)-2\Gamma(3/2)]\nonumber\\
&&-\frac{T^{3/2}}{\pi\sqrt{\Delta}}\Gamma(3/2)\zeta(1/2)+\cdots.
\eea
Remarkably, the $T^2$ term is cancelled out due to $\partial_\lambda F=0$ and the second line contributes 
a $F_T-F_0\propto T^{3/2}$, giving a $S=-dF/dT\sim\sqrt{T}$ entropy and a $\gamma=C/T\propto 1/\sqrt{T}$ specific heat coefficient.
\begin{figure}[h!]
\end{figure}
\subsection{Dynamical large-$N$ equations}
We start from the Hamiltonian (2) in the paper. The interaction part of the action is
\be
S_I=\frac{1}{\sqrt N}\sum_{ja\alpha}\intob{d\tau}[\chi_{ja}(\tau)\bar{b}_{j\alpha}(\tau)\psi\dn_{ja\alpha}(\tau)+{\rm h.c}]
\ee
Fist we integrate out the $c$-electron. The result is 
\bea
S_I&=&\frac{1}{N}\sum_{ja\alpha}\intob{d\tau_1d\tau_2}\Big[\chi_{ja}(\tau_1)b\dg_{j\alpha}(\tau_1)g_0(\tau_1-\tau_2)\nonumber\\
&&\hspace{5cm}b_{j\alpha}(\tau_2)\chi\dg_{ja}(\tau_2)\Big]\qquad,
\eea
where $g_{0} (\tau )$ is the bare local conduction electron propagator.
We decouple this by adding to the action a conjugate pair of two-point quantum fields
\be
S'=N\intob{d\tau_1d\tau_2}\hat{G}_B(\tau_2,\tau_1)\hat{\Sigma}_B(\tau_1,\tau_2)
\ee
by shifting
\bea
&&\hat G_B(\tau_2,\tau_1)\to \hat G_B(\tau_2,\tau_1)+\frac{1}{N}\sum_{j\alpha}b\dg_{j\alpha}(\tau_1)b\dn_{j\alpha}(\tau_2)\nonumber\\
&&\hat \Sigma_B(\tau_1,\tau_2)\to \hat \Sigma_B(\tau_1,\tau_2)+\frac{1}{N}g_0(\tau_1,\tau_2)\sum_{ja}\chi\dg_{ja}(\tau_2)\chi\dn_{ja}(\tau_1).\nonumber
\eea
we find
\bea
S'+S_I&\to& S'+\intob{d\tau_1d\tau_2}\Big[\sum_{n\alpha} b\dg_{n\alpha}(\tau_1)\hat\Sigma_B(\tau_1,\tau_2)b\dn_{n\alpha}(\tau_2)\nonumber\\
&&+\sum_{nk} \chi\dg_{nk}(\tau_1)g_0(\tau_2,\tau_1)\hat G_B(\tau_1,\tau_2)\chi\dn_{nk}(\tau_2)\Big].
\eea
The free energy (the effective action times $T$) is
\begin{eqnarray}\label{freen}
F[\hat G_B,\hat\Sigma_B]&=&NT{\rm Tr}\log[{\hat{\Sigma}_B(\tau_1,\tau_2)-G^{-1}_{B0}}(\tau_1,\tau_2)]-N\lambda s\nonumber\\
&-&KT{\rm Tr}\log[{g_0(\tau_2,\tau_1)\hat{G}_B(\tau_1,\tau_2)-G^{-1}_{\chi 0}(\tau_1,\tau_2)}]\nonumber\\
&+&NT\intob{d\tau_1d\tau_2}\hat{G}_B(\tau_2,\tau_1)\hat{\Sigma}_B(\tau_1,\tau_2)
\end{eqnarray}

where
\bea
&&G^{-1}_{B0}(\tau_1,\tau_2)=-(\partial_{\tau_1}+\lambda)\delta(\tau_1-\tau_2)\\
&&G^{-1}_{\chi 0}(\tau_1,\tau_2)=-J_K^{-1}\delta(\tau_1-\tau_2).
\eea
Here variables $\hat O$ with a hat on them, are fluctuating variables
that are integrated over inside the path integral. In
the limit of large-$N$, we can carry out a mean-field treatment of
the path integral by replacing it by its saddle-point value. Variation
of the free energy 
w.r.t. $\Sigma_B$ gives $[\Sigma_B-G_{B0}^{-1}]^{-1}+G_B=0$, so
$G_B$ and $\Sigma_B$ obey a Dyson equation.  
Before we
carry out the variation w.r.t $G_B$, it is convenient to define
$\Sigma_{\chi}(\tau_1,\tau_2)\equiv
g_0(\tau_2,\tau_1)G_B(\tau_1,\tau_2)$. If we furthermore define
$G_{\chi}(\tau_1,\tau_2)=[G_{\chi 0}^{-1}-\Sigma_{\chi}]\vert_{(\tau_1,\tau_2)}$, then
variation of the free energy w.r.t. $G_B$ gives
\begin{equation}\label{sigmab}
\Sigma_B(\tau_1,\tau_2)=-\gamma g_0(\tau_2,\tau_1)G_{\chi}(\tau_1,\tau_2)
\end{equation}
These set of self-consistent equations have a time-translationally
invariant solution, dependent only on the time difference
$\tau_1-\tau_2$. 

\begin{figure}
\includegraphics[width=\linewidth]{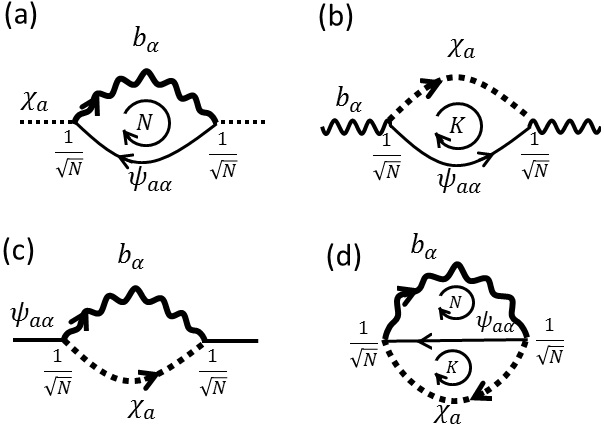}
\caption{(a-c) Diagrams for self-energy of (a) holon and (b) spinon and (c) the conduction electron. In two first two cases, the summation over the loop index (spin for holon and channel for boson) gives a factor of $N$ which compensates $1/N$ coming from vertices, whereas no compensation in (c) means that conduction electron self-energy vanishes to $O(1)$ and correspnding Green's function remains bare. (d) The interaction part of the Luttinger-Ward functional. The self-energies can be obtained by cutting the corresponding propagator in this diagram $\Sigma=\delta Y/\delta G$.}\label{figdiag}
\end{figure}

We now show that these mean-field equations can be obtained as the
saddle point of a Kadanoff-Baym free energy functional.
By identifying the argument of first logarithm in (\ref{freen})
as $G_{B}^{-1}$, we rewrite the free energy as 
\bean
(NL)^{-1}F[G_B]&=&T{\rm Tr}\Big\{\log[-G_B^{-1}]+(G_{B0}^{-1}-G_B^{-1})G_B\Big\}\\
&&-\lambda s+f_3[G_B], \\
f_3[G_B]&=&-\gamma T{\rm Tr}\log\{\Sigma_{\chi}[G_B]-G_{\chi 0}^{-1}\}
\eean
with $\Sigma_{\chi}[G_B]=g_0(\tau_2,\tau_1)G_B(\tau_1,\tau_2)$. If we take
variations of this expression w.r.t. $G_{B}$, we recover expression 
(\ref{sigmab}).  Next we elevate the free energy to a
functional of $G_{\chi }$  and $\Sigma_{\chi } $ by rewriting
$f_{3}$ as follows:
\bean
f_3[G_B]&\to& f_3[G_B,\Sigma_{\chi},G_{\chi}]\\
&=&-\gamma T{\rm Tr}\log[\Sigma_{\chi}-G_{\chi 0}^{-1}]-\gamma T\tr{\Sigma_{\chi} G_{\chi}}\\
&&+\gamma T\tr{\Sigma_{\chi}[G_B]\times G_{\chi}}.
\eean
Here, the last two terms basically cancel each other. 
If we set the variation of $f_{3}$ w.r.t. $G_{\chi}$ to zero, we
recover $\Sigma_{\chi}=\Sigma_{\chi}[G_B]$. 
If we set the variation of $f_{3}$ 
w.r.t $\Sigma_{\chi}$ 
to zero (only the first two terms have to be taken into account) 
we find $G_{\chi}=[G_{X_0}^{-1}-\Sigma_{\chi}]^{-1}$ as we had above.
Using the variation with respect to $\Sigma_{\chi}$ to get rid of $\Sigma_{\chi}$ in favor of $G_{\chi}$, we find
\bea
f_3&=&-\gamma T{\rm Tr}\Big\{\log[-G_{\chi}^{-1}]+(G_{\chi 0}^{-1}-G_{\chi}^{-1})G_{\chi}\Big\}\nonumber\\
&&+\gamma T\tr{\Sigma_{\chi}[G_B]\times G_{\chi}}
\eea
and this gives us final expression for the free energy in the form of a Luttinger-Ward functional of Green's functions
\bea
(NL)^{-1}F[G]&=&T\tr{\log(-G_B^{-1})+(G_{B0}^{-1}-G_B^{-1}) G_B}\nonumber\\
&&-\gamma T\tr{\log(-G_{\chi}^{-1})-(G_{\chi 0}^{-1}-G_{\chi}^{-1})G_{\chi}}\nonumber\\&&+T{\cal Y}[G_{\chi},G_B]-\lambda s,
\eea
where
\bea
{\cal Y}[G_{\chi},G_B]&=&\gamma\tr{\Sigma_{\chi}[G_B]\times G_{\chi}}\\
&=&\gamma\beta\intob{d\tau g_0(\tau)G_B(-\tau)G_{\chi}(\tau)}\equiv\beta Y\nonumber
\eea
or in real frequency
\bea
Y&=&\gamma\int{\frac{d\omega_1}{\pi}}\int\frac{d\omega_2}{\pi} f(\omega_1)f(\omega_2)\nonumber\\
&&\hspace{3cm}G_{\chi}''(\omega_1)\im{g_0(\omega_2)G_B(\omega_1+\omega_2)}\nonumber\\
&&-\gamma\int{\frac{d\omega_1}{\pi}}\int\frac{d\omega_2}{\pi}n_B(\omega_1)f(\omega_2)\nonumber\\
&&\hspace{3cm}G_B''(\omega_1)\im{g_0(\omega_2)G_{\chi}^*(\omega_2-\omega_1)}.\nonumber
\eea
The diagrammatic rationale for the self-energies (Eq.\,6 of manuscript) is shown in Fig.\,\pref{figdiag}. Note that since $\delta F[G]/\delta G=0$, the self-energies can be obtained from the Luttinger-Ward functional $\Sigma=\delta Y/\delta G$. $\Sigma_B(\tau)=-\gamma g_0(\tau)G_\chi(\tau)$ gives in real-frequency
\bea
\Sigma_B(\omega+i\eta)&=&\gamma\intinf{\frac{d\omega'}{\pi}}f(\omega')\Big\{g_c''(\omega')G^R_\chi(\omega-\omega')\nonumber\\
&&\hspace{2cm}-g^R_c(\omega+\omega')G_\chi''(-\omega')\Big\},\qquad\label{eqSigB}
\eea
and $\Sigma_\chi(\tau)=g_0(-\tau)G_B(\tau)$ gives in real-frequency
\bea
\Sigma_\chi^R(\omega+i\eta)&=&\intinf{\frac{d\omega'}{\pi}}\Big[-G_B^R(\omega+\omega')f(\omega')g_c''(\omega')\nonumber\qquad\\
&&\qquad+G_B''(\omega')n_B(\omega')g_c(\omega'-\omega-i\eta)\Big].\label{eqSigX}
\eea
To find self-consistent solution to the dynamical large-$N$ equations, we have implemented Eqs.\,\pref{eqSigB} and \pref{eqSigX} on a linear and logarithmic frequency grid, iteratively, together with the corresponding Dyson's equations. We start at high-temperature and gradually reduce the temperature to have convergence.

\begin{figure}
\includegraphics[width=\linewidth]{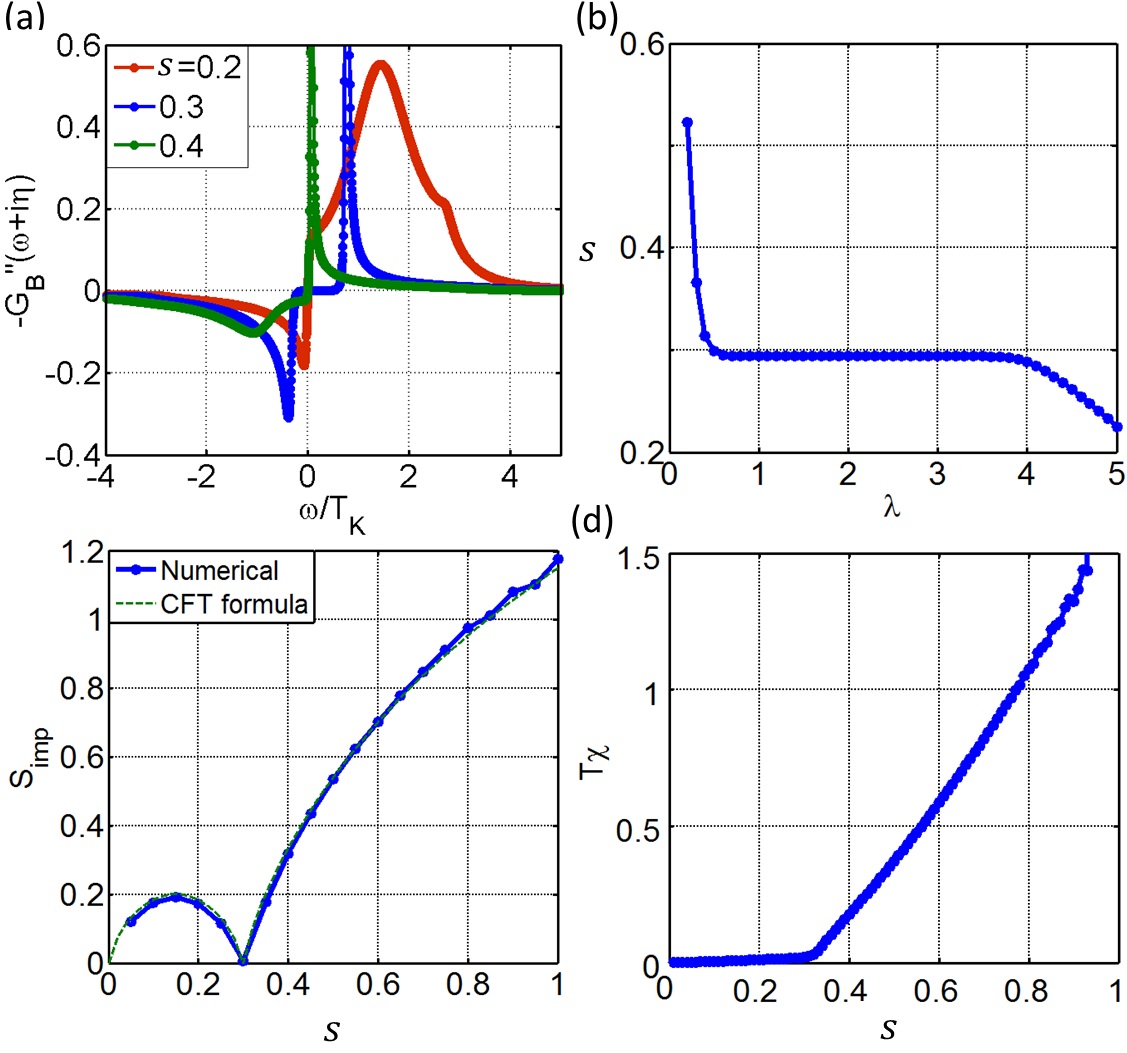}
\caption{The single-impurity multi-channel Kondo physics from the large-$N$ approach. $k=0.3$ is fixed and $s$ is varied. (a) The spectrum of spinons for fully-screened (blue) over-screened (red) and under-screened (green) cases. (b) The gap in $\lambda$ vs. $s$ for establizing a Fermi-liquid. (c) Residual entropy and (d) T$\chi$ as a function of $s$, which captures the Curie susceptibility coming from the residual unscreened moments.}\label{figimp}
\end{figure}

Fig.\,\pref{figimp} summarizes some aspects of the single-impurity Kondo physics as captured by the large-$N$ approach \cite{Parcollet1997}. 

\subsection{Entropy}
The free energy worked out in previous section is a stationary
functional with respect to $G_B$, $G_\chi$ and $\lambda$. Keeping those constant, we take derivative w.r.t $T$ to obtain the entropy. The result is \cite{Rech2006}
\bea
S (T)&=&-\int\frac{d\omega}{\pi}\partial_Tn_B(\omega)\Big\{{\rm Im}{\sum_k\log[-G_B^{-1}(k,\omega+i\eta)]}\nonumber\\
&&\hspace{3cm}+\Sigma''_B(\omega+i\eta)G'_B(\omega+i\eta)\Big\}\nonumber\\
&&-k\int\frac{d\omega}{\pi}\partial_Tf(\omega)\Big\{\im{\log[-G_{\chi}^{-1}(\omega+i\eta)]}\nonumber\\
&&\hspace{3cm}+\Sigma''_{\chi}(\omega+i\eta)G'_{\chi}(\omega+i\eta)\nonumber\\
&&\hspace{3cm}-g''_{0}(\omega+i\eta)\tilde\Sigma'_c(\omega+i\eta)\Big\}.\label{eq86}
\eea
Here, $g_0$ is the bare Green's function of conduction band. And
$\tilde\Sigma_c(\tau)=N\Sigma_c(\tau)$ where
$\Sigma_c(\tau)=\frac{1}{N}G_B(\tau)G_\chi(-\tau)$ is the self-energy
of the conduction electrons, which in the frequency domain is
\bea
\tilde\Sigma_C(\omega+i\eta)&=&\int{\frac{d\nu}{\pi}}\Big[n_B(\omega)G''_B(\nu+i\eta)G_\chi(\nu-\omega-i\eta)\qquad\\
&&\qquad -f(\nu)G''_\chi(\nu)G_B(\omega+\nu+i\eta)\Big].
\eea
At low temperature $\partial_Tf(\omega)\propto \beta\omega\delta'(\omega)$ and as the fermionic functions do not diverge in the large-$\Delta$ limit, the residual entropy is dominated by the bosonic terms in the first two lines of Eq.\,\pref{eq86}.
\subsection{Momentum sums}
In this section we sketch the derivation of Eq.\,\pref{eqGB} and the expression used for the calculation of the spin susceptibility.

According to Cauchy's theorem,
\be
F(z)=\oint{\frac{dz'}{2\pi i}}\frac{F(z')}{z'-z}=\int{\frac{d\omega}{\pi}}\frac{1}{\omega-z}\im{F(\omega+i\eta)}\label{eqCauchy}
\ee
where the contour in the first integral is counter-clockwise and in the second expression we have assumed that the the function $F(z')$ is analytical in the whole plane except on the real axis.

On the other hand for the local Green's function of the spinons we need to do the integral 
\be
G_{B}[\Omega(z)]=-\int{\frac{d\eps_{B}\rho(\eps_B)}{\eps_B-\Omega(z)}},\quad\rho(\eps_B)=\frac{(2\pi\Delta)^{-1}}{\sqrt{1-(\eps_B/2\Delta)^2}}.\nonumber
\ee
where $\Omega(z)=z-\lambda-\Sigma_B(z)$. The function $\rho_B(\eps_B)$ can be expressed as the imaginary part of an analytical function
\be
\rho(\eps_B)=-\im{F(\eps_B+i\eta)}, \quad F(z)=\frac{1}{\pi z\sqrt{1-(z/2\Delta)^{-2}}}.\nonumber
\ee
Therefore, using \pref{eqCauchy}, we find
\be
G_B[\Omega(z)]=\frac{1}{\Omega\sqrt{1-(\Omega/2\Delta)^{-2}}}.
\ee
The spin susceptibility can be expressed as the spinon bubble
\bea
&&\chi(k,\omega+i\eta)=\int{\frac{dq}{2\pi}}\int{\frac{d\omega'}{\pi}}n_B(\omega')G_B''(q,\omega')\\
&&\qquad[G_B(q+k,\omega'+\omega+i\eta)+G_B(q-k,\omega'-\omega-i\eta)].\nonumber
\eea
For the static zero momentum case $k=\omega=0$, we can write
\be
\chi=\int{\frac{d\omega}{\pi}n_B(\omega)}\im{\chi_\omega}, \quad \chi_\omega=\sum_kG_B^2(k,\omega+i\eta).
\ee
$\chi_\omega$ can be expressed as the derivative of $G_B(z)$,
\be
\chi_\omega=\int{\frac{d\eps_B\rho(\eps_B)}{(\eps_B-\Omega)^2}}=-\frac{d}{d\Omega}G_B(\omega+i\eta).
\ee
Therefore, we find
\be
\chi_\omega=\frac{1}{\Omega^2[1-(\Omega/2\Delta)^{-2}]^{3/2}}.
\ee
For the bosonic contribution to the entropy we need
\be
{\cal S}_\omega=\sum_k\log[-G_B^{-1}(k,\omega+i\eta)]=\int{d\eps_B\rho(\eps_B)\log[\eps_B-\Omega]}.\nonumber
\ee
Similar to above, ${\cal S}_\omega$ can be expressed as ${\cal S}_\omega\sim\int{d\Omega}G_B$. So,
\be
{\cal S}_\omega=\log\Big[1+\sqrt{1-(\Omega/2\Delta)^{-2}}\Big]+\log({\Omega}/2)\nonumber
\ee
assuming $\re{\Omega/2\Delta}\le -1$.
\subsection{Applying magnetic field}

We assume that the $B$-field couples to all but one boson $b_1$
increasing their energy, so that the field-dependent ``Zeeman'' term
in the Hamiltonian takes the form
\[
H_{Z}= B\sum_{\alpha \neq 1}b\dg_{\alpha }b_{\alpha }
\]
 For an isolated spin, this means that one of
the bosons has the energy $\lambda$ whereas the others have the energy
$\lambda+B$. There are two phases \cite{Coleman2003,Coleman2005a}: a
paramagnetic phase at high temperature in which the population of the
low-energy boson is negligible $\braket{b_1}=0$ and $\lambda$ is
adjusted so that $n_B(\lambda+B)=s$. Lowering the temperature
$\lambda$ decreases until it becomes zero at $T/B=1/\log[1+1/s]$ below
which the low-energy boson undergoes a Bose-Einstein condensation
(BEC) $\braket{b_1}^2=s-n_B(B)$ to accommodate an $O(N)$
magnetization, the polarized phase. In presence of Kondo screening and
the spinon hopping, essentially similar arguments apply except that
the spinon energies are dressed by the hopping and renormalized by the
spin fluctuations. We start by writing the Hamiltonian in a magnetic
field ($L$ is the system size) 
\bea
H&=&N\sum_{ja}\frac{\chi\dg_{ja} \chi\dn_{ja}}{J_K}+\sum_{ja\alpha}(\chi\dn_{ja}b\dg_{j\alpha} \psi\dn_{ja\alpha}+{\rm h.c})\\
&+&\lambda\sum_j[\sum_{\alpha}b\dg_{j\alpha} b\dn_{j\alpha}-2S]+B\sum_{j,\alpha\neq 1}b\dg_{j\alpha} b\dn_{j\alpha}+H_C\nonumber\\
&-&{\Delta}\sum_{j\alpha}[b\dg_{j\alpha}b\dn_{j+1,\alpha}+{\rm h.c}]+N\frac{\Delta^2}{J_H}.
\eea
Separating the low-energy boson from the rest we have
\bea
H&=&N\sum_{ja}\frac{\chi\dg_{ja} \chi\dn_{ja}}{J_K}+\sum_{ja,\alpha\neq 1}(\chi\dn_{ja}b\dg_{j\alpha} \psi\dn_{ja\alpha}+{\rm h.c})\nonumber\\
&+&N\frac{\Delta^2}{J_H}-{\Delta}\sum_{j,\alpha \neq 1}[b\dg_{j\alpha}b\dn_{j+1,\alpha}+{\rm h.c}]\nonumber\\
&+&(\lambda+B)\sum_j[\sum_{j,\alpha\neq 1}b\dg_{j\alpha} b\dn_{j\alpha}-2S]+2SB+H_C\nonumber\\
&+&({\lambda-2\Delta})\sum_jb\dg_{j,1}b\dn_{j,1}+\sum_{ja}(b\dg_{j,1}\chi_{ja} \psi\dn_{ja,1}+{\rm h.c}).\qquad
\eea
We consider a mean-field solution in which the low-energy boson condenses $b_1\to\braket{b_1}$. The condition for BEC is that the energy of $b_1$ boson becomes zero. But the apparent energy $\lambda-2\Delta$ is further renormalized by the charge fluctuations in the last term. After the condensation, the Hamiltonian is (by $\chi\to \tilde\chi/\sqrt{N}$)
\bea
H&=&\sum_j\frac{\tilde\chi\dg_{ja} \tilde\chi\dn_{ja}}{J_K}+\frac{1}{\sqrt N}\sum_{ja,\alpha\neq 1}(\tilde\chi\dn_{ja}b\dg_{j\alpha} \psi\dn_{ja\alpha}+{\rm h.c})\nonumber\\
&-&{\Delta}\sum_{n,\alpha \neq 1}[b\dg_{n\alpha}b\dn_{n+1,\alpha}+{\rm h.c}]+N\frac{\Delta^2}{J_H}+H_C\nonumber\\
&+&N(\lambda+B)[\frac{1}{N}\sum_{\alpha\neq 1}b\dg_\alpha b\dn_\alpha-q]+NBq\nonumber\\
&+&NL(\lambda-2\Delta)\braket{b_1}'^2+\braket{b_1}'\sum_{ja}(\tilde\chi_{ja} \psi_{ja,1}+{\rm h.c})\qquad
\eea
Since, $\braket{b_1}$ is extensive, we have defined $\braket{b_1}'=\braket{b_1}/{\sqrt N}$. As long as the $\braket{b_1}'=0$, $\lambda$ adjusts itself so that the spectrum does not move. Also, note that once $b_1$ condenses in a magnetic field, the value of $\Delta$ might change, but for small $B$ this is negligible and we have discarded the $B$-dependence of the mean-field $\Delta$ in this paper. To find the condensate fraction, we minimize the energy with respect to $\braket{b_1}'$. 
So we find
\bea
(\lambda-2\Delta)\bar b_1'&=&\frac{1}{NL}\sum_{ja}{\rm Re}\braket{\psi_{ja,1}\chi_{ja}}
\eea
Considering that
\be
\lim_{t\to 0}\braket{\psi_{ja,1}(t)\chi_j(0)}=
i\int\frac{d\omega}{2\pi}G_{\psi\chi }^<(\omega),
\ee
and the relation
\be
G^<_{c \chi }(\omega)=-f(\omega)[{G_{\psi\chi }(\omega+i\eta)-G_{\psi\chi }(\omega-i\eta)}],
\ee
we can write
\bea
(\lambda-2\Delta)\bar b'_1&=&\frac{1}{NL}{\rm Re}\sum_j\lim_{t\to 0}\braket{\psi_{ja,1}(t)\chi_{ja}(0)}\label{eq39}\\
&&\hspace{-1cm}=\gamma\int\frac{d\omega}{\pi}f(\omega)\im{G_{\psi\chi}(\omega+i\eta)-G_{\psi\chi}(\omega-i\eta)}.\nonumber
\eea
We can use equation of motion (EoM) to calculate the mixed function but first, better write $\psi_{ja,1}$ in momentum space:
\be
\hspace{-0.2cm}G_{\psi\chi}(\tau)\equiv\braket{-Tc_{ja,1}(\tau)\chi_{ja}}=\sum_k\braket{-Tc_{ja,1k}(\tau)\chi_{ja}}.
\ee
Note that $c_{ja,1k}$ refers to $\alpha=1$ and $k$ is the electron momentum. EoM gives
\bea
-\partial_\tau G_{c\chi}(\tau)&=&\braket{-T[c_{ja1,k},H]_\tau \chi_{ja}}\nonumber\\
&=&\braket{-T\{\eps_k c_{ja1,k}(\tau)+\bar b_1'\chi\dg_{ja}(\tau)\} \chi_{ja}},
\eea
or
\be
(-\partial_\tau-\eps_k)G_{c\chi}(\tau)=\bar b_1'\braket{-T\chi\dg_{ja}(\tau)\chi_{ja}},
\ee
which after momentum summation leads to
\be
G_{\psi\chi}(\tau)=b_1'\sum_k g_k(\tau)* \braket{-T\chi\dg_{ja}(\tau)\chi_{ja}}.
\ee
To find the Fourier transform of the last term,
note that fermionic Lehmann representation of $\langle{-T\chi\dn_j(\tau)\chi_j\dg}\rangle$ is
\be
G(i\omega_n)=-\frac{1}{Z}\sum_{mn}\abs{\braket{n\vert\chi\vert m}}^2\frac{e^{-\beta E_n}+e^{-\beta E_m}}{i\omega_n+E_n-E_m}
\ee
Therefore, $\chi\to\chi\dg$ corresponds to $n\lr m$ and $G(i\omega_n)\to-G(-i\omega_n)$. So,
\be
G_{\psi\chi}^R(\omega)=-\bar b_1' g^R(\omega)G_{\chi }^A(-\omega).
\ee
Plugging this into Eq.\,\pref{eq39} we find
\be
\lambda-2\Delta=-\gamma \int{\frac{d\omega}{\pi}}f(\omega)[g''_{c1}(\omega)G'_\chi(-\omega)-g'_{c1}(\omega)G''_\chi(-\omega)],
\ee
and of course $n_B(\lambda)=s-m$ where we defined the
magnetization, $m \equiv {\bar b}_1'^2$. In the case of an isolated spin, it condensed whenever its energy $\lambda=0$ becomes zero. The above equation is generalization of that formula to the Kondo case. The self-energy of $\chi$ is modified
\be
\Sigma_\chi(\tau)=\frac{N-1}{N}g_c(-\tau)G_{B,\alpha\neq 1}(\tau)-\bar b_1'^2g_{c}(-\tau),
\ee
so that in large-$N$ we have
\be
\Sigma_\chi(\omega+i\eta)=\Sigma_\chi^{\rm old}(\omega+i\eta)-m g_c(-\omega-i\eta)
\ee
and the self-energy of the bosons is not modified. \\

\subsection{Thermodynamical properties in presence of magnetic field}
\begin{figure}[h!]
\includegraphics[width=\linewidth]{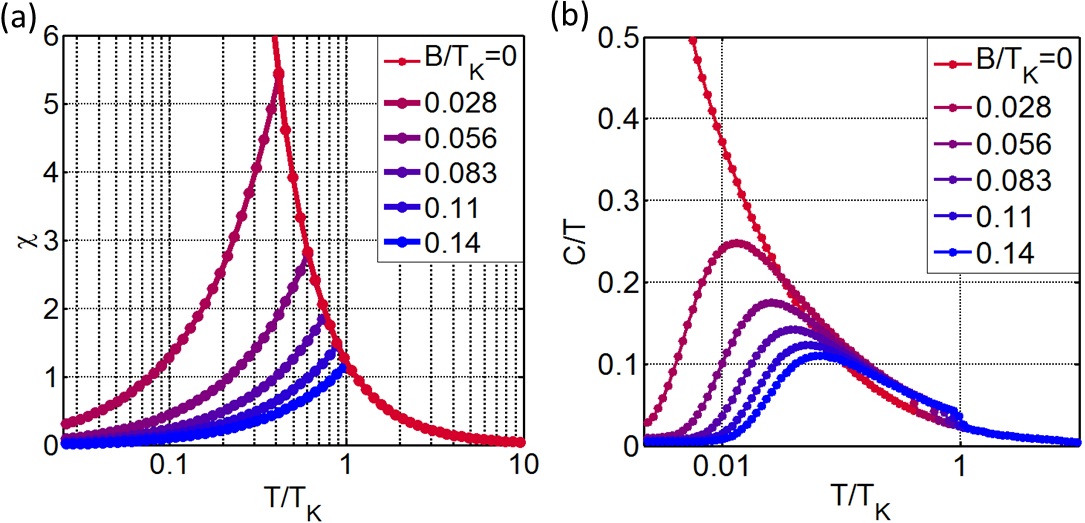}
\caption{Application of a magnetic field at $T_K/J_H=0.1$ drives the
(a) susceptibility $\chi$ and (b) specific heat coefficient $\gamma$
from the critical behavior (red) to a FL behavior (blue). The kinks
are due to a finite temperature FM
transition induced by the field.}\label{fig4}
\end{figure}

We have computed the spin susceptibility and specific heat in presence of a magnetic field using the modification discussed in last section. The result is shown in Fig.\,\pref{fig4}. The Fermi liquid is robust against a magnetic field, and a field produces small quadratic shifts in the various mean-field
quantities. 
However, in the gapless phases, application of a small magnetic field \cite{Coleman2003,Coleman2005a} has a profound effect: it reinstates 
Fermi-liquid behavior with an scale $T_B$ set by the Zeeman energy (at the QCP) or a combination of the spinon
bandwidth and magnetic field (in the FM phase). 

\end{document}